\documentclass{article}

\usepackage{arxiv}

\usepackage[utf8]{inputenc} 
\usepackage[T1]{fontenc}    
\usepackage{hyperref}       
\usepackage{url}            
\usepackage{booktabs}       
\usepackage{amsfonts}       
\usepackage{nicefrac}       
\usepackage{microtype}      
\usepackage{lipsum}		
\usepackage{graphicx}
\usepackage{doi}
\usepackage{multirow}
\usepackage{amsmath}
\usepackage{algorithm}
\usepackage[numbers]{natbib}

\title{Optimized two-stage AI-based Neural Decoding for Enhanced Visual Stimulus Reconstruction from fMRI Data}

\author{ \href{https://orcid.org/0000-0001-9204-3363}{\includegraphics[scale=0.06]{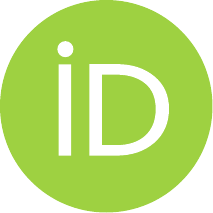}\hspace{1mm}Lorenzo Veronese} \\
	Department of Electronics,\\
	Information, and Bioengineering,\\
	Politecnico di Milano, Milan, Italy. \\
	\texttt{lorenzo.veronese@polimi.it} \\
	\And
    \href{https://orcid.org/0000-0002-3365-580X}
 {\includegraphics[scale=0.06]{orcid.pdf}\hspace{0.1mm}Andrea Moglia} \\
	Department of Electronics,\\
	Information, and Bioengineering,\\
	Politecnico di Milano, Milan, Italy. \\
 \And
	\href{https://orcid.org/0000-0002-6276-6314}
 {\includegraphics[scale=0.06]{orcid.pdf}\hspace{0.1mm}Luca Mainardi} \\
	Department of Electronics,\\
	Information, and Bioengineering,\\
	Politecnico di Milano, Milan, Italy. \\
    \And
	\href{https://orcid.org/0000-0003-3995-8673}
 {\includegraphics[scale=0.06]{orcid.pdf}\hspace{0.1mm}Pietro Cerveri} \\
	Department of Electronics,\\
	Information, and Bioengineering\\
	Politecnico di Milano, Milan, Italy. \\
        Università di Pavia, Pavia, Italy.
}

\date{}

\hypersetup{
pdftitle={Cascade learning in multi-task encoder-decoder networks for concurrent bone segmentation and glenohumeral joint assessment in shoulder CT scans},
pdfsubject={cs.CV, cs.AI},
pdfauthor={Luca Marsilio, Davide Marzorati, Matteo Rossi, Andrea Moglia, Luca Mainardi, Alfonso Manzotti, Pietro Cerveri},
pdfkeywords={CT segmentation, Glenohumeral joint diagnostic, Deep learning, PSI-based intervention, Shoulder arthroplsty},
}

\begin{document}
\maketitle


\begin{abstract}
AI-based neural decoding reconstructs visual perception by leveraging generative models to map brain activity, measured through functional MRI (fMRI), into latent hierarchical representations. Traditionally, ridge linear models transform fMRI into a latent space, which is then decoded using latent diffusion models (LDM) via a pre-trained variational autoencoder (VAE). Due to the complexity and noisiness of fMRI data, newer approaches split the reconstruction into two sequential steps, the first one providing a rough visual approximation, the second on improving the stimulus prediction via LDM endowed by CLIP embeddings. This work proposes a non-linear deep network to improve fMRI latent space representation, optimizing the dimensionality alike.  
Experiments on the Natural Scenes Dataset showed that the proposed architecture improved the structural similarity of the reconstructed image by about 2\% with respect to the state-of-the-art model, based on ridge linear transform. The reconstructed image's semantics improved by about 4\%, measured by perceptual similarity, with respect to the state-of-the-art. The noise sensitivity analysis of the LDM showed that the role of the first stage was fundamental to predict the stimulus featuring high structural similarity. Conversely, providing a large noise stimulus affected less the semantics of the predicted stimulus, while the structural similarity between the ground truth and predicted stimulus was very poor. The findings underscore the importance of leveraging non-linear relationships between BOLD signal and the latent representation and two-stage generative AI for optimizing the fidelity of reconstructed visual stimuli from noisy fMRI data.
\end{abstract}


\section{Introduction}
\label{sec:introduction}
Understanding the intricate relationship between neural activity and visual perception is one of the central aims in neuroscience research~\cite{Kamitani2005, Zafar2015}. Functional magnetic resonance imaging (fMRI) has been described to offer a window into such a relationship by capturing blood-oxygen-level-dependent (BOLD) signals that indirectly reflect neuron functional patterns  \cite{Cox2003}. However, the complex and noisy nature of fMRI data presents a significant challenge in reconstructing the underlying visual stimuli. Likewise, significant barriers to accurate reconstruction have been described to emerge when dealing with a scene conveying rich semantic meaning, compared to reconstructing low-complexity content~\cite{ Miyawaki2008}. Leveraging the power of deep neural networks (DNNs), novel image decoding approaches have been pioneering to shed light on how humans learn to decode and interpret visual information \cite{Zafar2015, Shen2019}. For instance, deep networks have been very recently investigated to emulate perceptual systems for tasks like object recognition \cite{Li2024}, mimic gradual learning in categorization taksk\cite{Davidson2024}, and even simulate how early visual abilities can adapt as the brain matures \cite{Mathis2024}. In the field of multiple hierarchical representations, based on DNNs, the underlying idea is that the BOLD signal (multi-voxel fMRI activation pattern) is implicitly conveying latent information. This information can be decoded using a similar approach to how features are extracted from images in deep convolutional neural networks, like the well-known ImageNet model~\cite{Shen2019}. 
The BOLD signal is typically preprocessed using General Linear Model (GLM) analysis, which generates single-trial response estimates, known as betas, for each voxel in response to individual stimuli. Subsequently, the relationship between the beta weights and the derived latent space is established using techniques such as ridge linear regression~\cite{NunezElizalde2019}. The relationship between the original fMRI BOLD space, which contains the raw activation patterns across voxels (brain regions), and the derived latent spaces is established through techniques like ridge linear regression~\cite{NunezElizalde2019}. This method builds upon ordinary least squares regression, but with an added regularization term that helps prevent overfitting~\cite{DuprelaTour2022}. By applying ridge regression (or similar techniques), the complex BOLD signal is essentially compressed into a lower-dimensional latent space.
This compressed signal then becomes the key for decoding. Various deep learning architectures, like Variational Autoencoders (VAE)~\cite{Sun2023}, 
Generative Adversarial Networks (GAN) ~\cite{Wang2022}, 
and more recently Latent Diffusion Models (LDM)~\cite{Takagi2023, Ferrante2024}, 
have been employed for this purpose. While spatially related to the original visual stimulus, the reconstruction from the BOLD latent signal was a very coarse low-resolution image~\cite{Shen2019}. Increasing the size of the latent space up to 30 layers was documented to be very computationally demanding ~\cite{Ozcelik2023}, providing nonetheless questionable advantage in the reconstruction quality ~\cite{Scotti2023}.
In order to empower the reconstruction ability of the visual stimulus, a second processing stage has been recently proposed by embedding Contrastive Language-Image Pre-training (CLIP) models~\cite{Lin2024, Meng2023, Gu2024}. CLIP models are pre-trained on a massive dataset (400 million) of image-text pairs available on the Internet, allowing them to learn the relationship between visual and language features. Hence using CLIP, decoding models exploit concurrently BOLD and text latent representations. This enables the link of specific brain activity patterns to visual elements described in the text, which has been documented to improve the reconstruction quality~\cite{Scotti2023, Ozcelik2023}. In order to enable the use of such information, the relation between CLIP space and BOLD latent space must be learned, for instance using ridge linear regression models as in the first stage. However, how output uncertainty of the first processing stage affected the second stage was not systematically investigated, posing questions about the real utility of the first processing stage.  
In this work, we mainly focused on three main shortcomings, namely the linear mapping of the BOLD signal to the latent space, the large dimensional VAE space, and the sensitivity of the final reconstruction to the uncertainty of the first stage output. Despite the traditional ridge linear regression, we here proposed to map the BOLD signal in the latent representation by means of a non-linear neural network. In order to cope with the computational demands of the VAE dimensionality, an ablation analysis was carried out leading to optimizing the BOLD encoding size.  
The effectiveness of the improvements and analyses was tested using the Natural Scenes Dataset (NSD) \cite{Allen2021} acquired with 7T fMRI measurements on eight healthy subjects while they visualized natural color images from the Common Objects in Context
(COCO) dataset, composed of 73,000 images \cite{Lin2014a}.
Our main contributions can be thus synthesizes as follows: 
\begin{itemize}
    \item modeling latent space of beta weights using gated-recurrent-unit (GRU) network, which was tested against other deep architecture and state-of-the-art linear ridge regressor;
    \item optimization of the latent space size with respect to output quality and computational cost;
    \item noise sensitivity analysis to increase the understanding of the first stage role in the final stimulus reconstruction.  
\end{itemize}

\section{Related Works}

\subsection{Multimodal Contrastive Learning Pre-training Models}
The CLIP model \cite{Rombach2022} is a special multi-modal neural network devoted to building a shared embedding space linking images and textual descriptions. A CLIP model is usually pre-trained on a very large amount of image-text pairs, using contrastive loss functions \cite{Zhang2020}. For instance, the symmetric cross-entropy loss maximizes the cosine similarity of image-text pairs that are true matches and minimizes it for mismatching pairs as: 
\begin{equation}
L =\sum_{i=1}^N \log \frac{\exp(\text{sim}(\mathbf{z}_{I_i}, \mathbf{z}_{T_i})/\tau)}{2N\sum_{j=1}^N \exp(\text{sim}(\mathbf{z}_{I_i}, \mathbf{z}_{T_j})/\tau)} + \sum_{i=1}^N \log \frac{\exp(\text{sim}(\mathbf{z}_{T_i}, \mathbf{z}_{I_i})/\tau)}{2N\sum_{j=1}^N \exp(\text{sim}(\mathbf{z}_{T_i}, \mathbf{z}_{I_j})/\tau)}
\end{equation}

\noindent where $N$ is the image-text pair batch size, $\text{sim}(\mathbf{z}_{1}, \mathbf{z}_{2})$ denotes the cosine similarity between the CLIP embedding ${z}_{1}$ and ${z}_{2}$, while $\tau$ is a temperature parameter on the logits.
This procedure enables the model to perform zero-shot transfer for downstream tasks such as classification, where the model is capable of categorizing images without having seen any examples of the categories during training. Thanks to its versatility, it was proposed in a variety of tasks, from image generation from captions~\cite{Ramesh2022}, to multi-context blending~\cite{Xu2023}, to content retrieval~\cite{PortilloQuintero2021}, to, as in our case, neural decoding from both fMRI and electroencephalogram (EEG)~\cite{Ferrante2024}. In the context of neural decoding from fMRI signals, CLIP embeddings are often employed to guide Conditioned GANs \cite{Ozcelik2022} or Latent Diffusion Models (LDMs) \cite{Ozcelik2023}. These embeddings, predicted from the beta signals, facilitate the reconstruction process by providing a predominantly semantic framework. The incorporation of CLIP embeddings enhances the semantic coherence over purely pixel-level accuracy, not only advancing the fidelity of the reconstructions but also providing insights into the neural representation of semantic information in the brain.

\subsection{Image decoding by Latent Diffusion Models}
Latent diffusion models (LDMs)~\cite{Ho2020} have emerged as powerful tools for image synthesis across various domains. Despite traditional diffusion models that progressively add noise to an image, LDMs operate on a latent space~\cite{Wang2024}, involving three main steps. The first step consists of encoding the input image (in the training phase, it is the ground truth image) into a low-dimensional latent space:
\begin{equation}
    \mathbf{z} = E(\mathbf{x})
\end{equation}
where $E$ is the encoding model that produces the compressed, latent representation $z$ of the image $x$. A Gaussian noise is then iteratively added over $T$ time steps. This is called the forward diffusion process. Each addition of noise is regulated by the following equation:
\begin{equation}
    q(\mathbf{z}_t | \mathbf{z}_{t-1}) = \mathcal{N}(\mathbf{z}_t; \sqrt{1-\beta_t}\mathbf{z}_{t-1}, \beta_t \mathbf{I})
\end{equation}
where $\beta_t$ is a variance schedule determining the amount of noise added at each step $t$. This process produces a set of images (in the latent space) that are one noisier than the other, progressively. Then, a denoiser model, usually a convolutional neural network, such as the UNet~\cite{Ronneberger2015}, is instantiated in order to learn the reverse mapping that removes the noise step by step. Hence, the model learns to perform the so-called reverse diffusion process as: 
\begin{equation}
    p_\theta(\mathbf{z}_{t-1} | \mathbf{z}_t) = \mathcal{N}(\mathbf{z}_{t-1}; \mu_\theta(\mathbf{z}_t, t), \sigma^2_t \mathbf{I}),
\end{equation}
where \( \mu_\theta \) and \( \sigma^2_t \) are parameters learned to approximate the mean and variance of the reverse process. The objective function typically used is:
\begin{equation}
    L(\theta) = \mathbb{E}_{t, z_0, \epsilon, y} \left[ \| \epsilon - \epsilon_\theta (z_t, t, \tau_\theta (y)) \|^2 \right]
\end{equation}
where $\epsilon$ is the ground truth Gaussian noise, $\epsilon_\theta$ is the noise predicted by the UNet from latent variable $z_t$, at time $t$, conditioned on $\tau_\theta (y)$ (e.g. the CLIP embeddings). The reverse diffusion reconstructs the final generated image $\widetilde{x}$ as:
\begin{equation}
    \mathbf{\widetilde{x}} = D(\mathbf{z})
\end{equation}

\noindent where $D$ is the decoder model and $z$ is the generated latent representation.
Leveraging Latent Diffusion Models (LDMs) for neural decoding enables operations within a latent space that captures compressed, abstract features of images, rather than focusing on pixel-level details. Combined with guidance from CLIP embeddings, this approach integrates semantic information into the reconstruction process. Moreover, the latent space of LDMs aligns closely with the brain's hierarchical representations, spanning from low-level visual features (e.g., edges, textures) to high-level semantic concepts. This alignment establishes a compelling parallel between the neural decoding architecture and the underlying brain signals, enhancing both interpretability and fidelity in the reconstructions.

\section{Methodology}
\subsection{Getting beta weights from BOLD signal}
\label{getting_beta_weights_from_bold_signal}
The hemodynamic response function (HRF) describes how blood flow changes in response to neural activity, a process central to interpreting brain imaging, especially fMRI. When a region of the brain activates, neurons there consume more oxygen, leading to local increases in blood flow to meet the demand. This response forms the basis of the blood-oxygen-level-dependent (BOLD) signal. Event-related acquisition designs are characterized by the rapid succession of visual stimuli, triggering HRF with a corresponding increase in the BOLD signal shortly after image presentation. Typical HRFs last for about 30 seconds. In fast event-related designs, where interstimulus intervals are only a few seconds, this leads to the overlap of multiple HRFs during acquisition sessions, causing the effects of different stimuli to blend at each time point. A beta value quantifies the extent to which a voxel is activated in response to a single stimulus. In the present study, beta values were derived exploiting generalized linear models, implemented in the public library GLMsingle package ~\cite{Prince2022}, using the following voxel-wise equation~\cite{Kay2013}:
\begin{equation*}
    \mathbf{y} = (\mathbf{k}\mathbf{X}) \mathbf{\beta} + \mathbf{P} \mathbf{u} + \mathbf{G} \mathbf{v} + \mathbf{\epsilon} 
\end{equation*}

where $\textbf{y}$ represents the fMRI time series, $\boldsymbol{X}$ is a design matrix indicating the presence of each visual stimulus at corresponding timestamps, $\boldsymbol{k}$ is the discretized fixed HRF, $\boldsymbol{\beta}$ denotes the vector of beta weights, and $\boldsymbol{\epsilon}$ symbolizes the noise term. Additionally, $\boldsymbol{P}$ and $\boldsymbol{G}$ represent the polynomial and noise regressors, respectively, while $\boldsymbol{u}$ and $\boldsymbol{v}$ are their associated weights. The polynomial regressors modelled the baseline signal level, while the noise regressors accounted for signal fluctuations unrelated to the experimental conditions. Beta weights were iteratively computed using Ordinary Least Squares (OLS) as follows:
\begin{equation}
   \begin{bmatrix} \boldsymbol{\beta} \\ \boldsymbol{u} \end{bmatrix} = \left( \langle \begin{bmatrix} \boldsymbol{kX} &  \boldsymbol{P} \end{bmatrix},\begin{bmatrix} \boldsymbol{kX} &  \boldsymbol{P} \end{bmatrix}\rangle \right)^{-1} \begin{bmatrix} \boldsymbol{kX} &  \boldsymbol{P} \end{bmatrix}
\end{equation}

\noindent where $\boldsymbol{G}$ form factor was computed separately using principal component analysis. The overall procedure incorporated ridge regression regularization~\cite{Rokem2020} to further mitigate noise components, as well as a library of 20 candidate HRFs ($\boldsymbol{k}$) to evaluate the best fit (Algorithm \ref{alg:glmsingle_algorithm}).

\begin{algorithm}
\caption{Beta estimation using GLMsingle library}
\label{alg:glmsingle_algorithm}
    \textbf{Inputs}: HRF library, BOLD volumes, design matrix
    
    \textbf{Outputs}: Single-trial betas estimates
    
    \textbf{Step 1}: Estimate HRF by iterative fitting

    \hspace{1em} \textbf{1.1}: Set one available HRF

    \hspace{1em} \textbf{1.2}: Estimate $\boldsymbol{\beta}$ and $\boldsymbol{u}$ using OLS

    \hspace{1em} \textbf{1.3}: Repeat 1.1 and 1.2 over the 20 HRFs of the library
    
    \textbf{Step 2}: Denoising
    
    \hspace{1em} \textbf{2.1}: Compute voxel-wise cross-validated $R^2$
    
    \hspace{1em} \textbf{2.2}: Select voxels with $R^2<0$ as noise pool
    
    \hspace{1em} \textbf{2.3}: Compute noise regressors using PCA on timeseries of the noise pool
    
    \hspace{1em} \textbf{2.4}: Select number of noise regressors through 
    
    \hspace{2.8em} cross-validation

    \textbf{Step 3}: Ridge regularization
\end{algorithm}

\subsection{Natural Scenes Dataset}
\begin{figure} [t]
    \centering
    \includegraphics[scale=1]{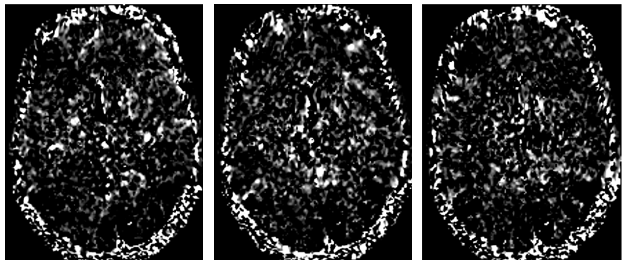}
    \caption{Examples of a transverse section beta coefficient images for subject \#1 in the NSD dataset. These volumes were acquired while watching samples 1, 6, and 8 (from left to right) of the test set. For each volume, the transverse section at voxel 80 was considered and values below the $40^{th}$ percentile and above the $95^{th}$ percentiles were saturated to pure black and white respectively.}
    \label{fig:example_fmri}
\end{figure}
The fMRI-stimulus pairs were obtained by the open-source Natural Scenes Dataset (NSD) ~\cite{Allen2021}. This comprised 7T fMRI recordings, along with BOLD signals, from eight healthy participants. During continuous acquisitions, they viewed natural color images, taken from the Common Objects in Context (COCO) dataset \cite{Lin2014a}. COCO images were 425×425 pixels and accompanied by a description in English natural language. The beta coefficients were computed using data in the NSD (Fig. \ref{fig:example_fmri}), according to the methodology described in sec. \ref{getting_beta_weights_from_bold_signal}) . 
In NSD, pre-defined ROIs were available for each subject, allowing to focus on relevant voxels and distinguish between low-level visual areas (such as V1, V2, V3, and V4) and higher-order visual areas involved in recognizing faces, places, words, and body parts within the inferior temporal cortex. Z-score normalization was applied to beta values, multiplying each sample by the mean and dividing by its standard deviation, both computed on the entire dataset. According to the literature \cite{Scotti2023, Ozcelik2023}, 90\% of the samples was used to build the training set, while the remaining 10\% composed the test set. Out of the eight participants, just four (\#1, \#2, \#5, and \#7) were retained for this study because they underwent the full acquisition protocol. As an example, for subject \#1, 37 acquisition sessions were performed, each acquisition involving 750 stimulus images (administration time 4~s). For each stimulus, the BOLD signal was acquired in that time frame. General model transform mapped the BOLD signal into the beta values. After applying the above-cited ROIs, 9,841 one-dimensional vectors of beta values were attained, each composed of 15,724 elements.

\subsection{Neural decoding architecture and training}
\begin{figure}[t]
    \centering
    \includegraphics[width=0.97\linewidth]{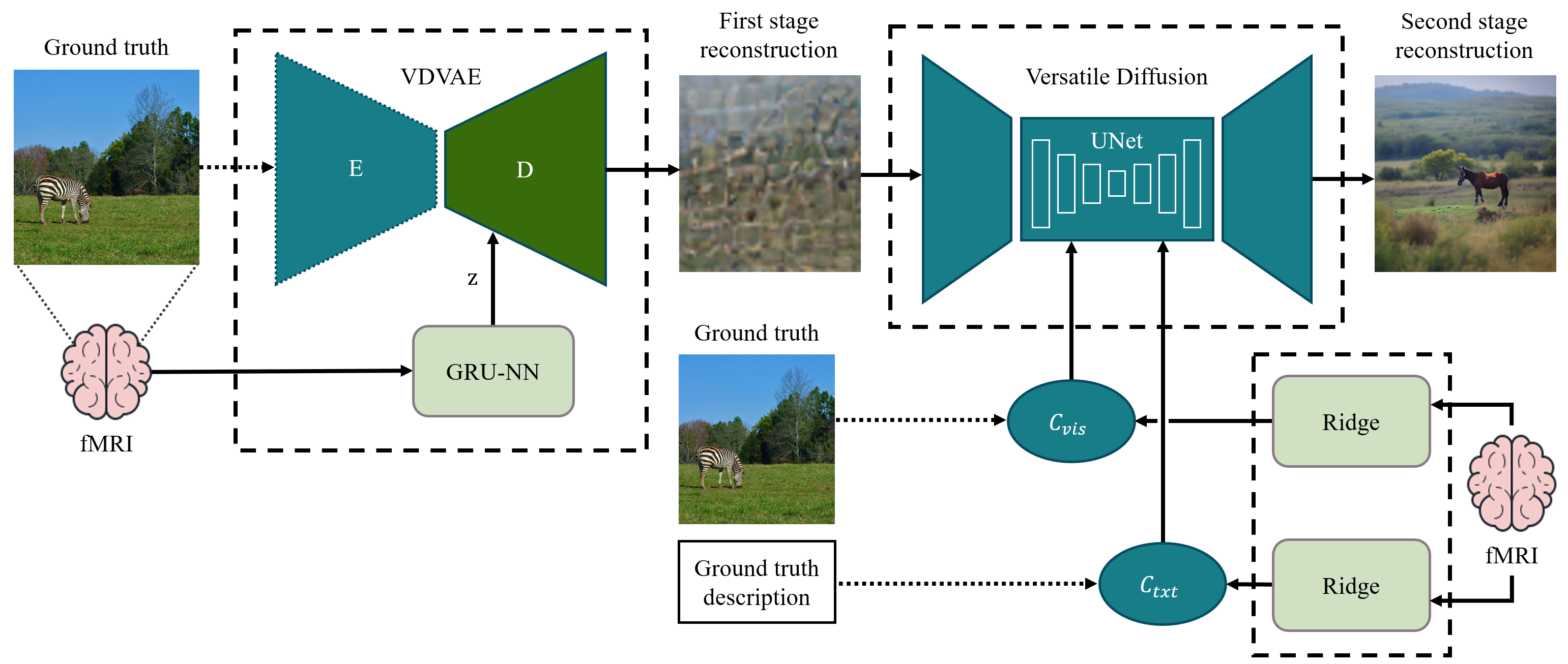}
    \caption{Model architecture.  In the I stage, the GRU-based network process the fMRI data (beta coefficients) to predicts the latent variables of the VDVAE decoder. The VDVAE predicts a rough image reconstruction that becomes the input of the II stage. The LDM, performing a reverse diffusion process, is conditioned with CLIP-vision and CLIP-text embeddings to predict the final image reconstruction. Dashed lines indicate elements present only in the training step.}
    \label{fig:architecture}
\end{figure}

The reconstruction process unfolded in two sequential steps. 
First, the beta images were processed by the proposed GRU-based neural network, producing the latent variables of the decoder of a Very-Deep Variational Autoencoder (VDVAE) ~\cite{Child2020} (Fig. \ref{fig:architecture}). The output of the VDVAE, namely the visual reconstruction of the stimulus, was a spatially organized but blurred initial guess. The second stage, composed of a latent diffusion model (LDM), took in input the initial guess and estimated in the output a refined reconstruction of the stimulus. The LDM comprised an autoencoder front-end, an inner UNet-like  model~\cite{Ronneberger2015}, working in the image and text latent spaces, and an output decoder (Fig. \ref{fig:architecture}). The role of the inner UNet model was to perform progressive denoising, being guided by CLIP embeddings of text and vision simultaneously, which were injected into the UNet through cross-attention mechanism~\cite{Socher2013}. These two latent representations were predicted by linear regressors utilizing the original $\beta$ signal. The final output was a high-resolution image exhibiting semantic coherence with the ground truth image.

\subsubsection{First stage: GRU-based network training}
\label{subsec:first_step_VDVAE_reconstruction}
This step aimed at reconstructing a first approximation of the image, exploiting the VDVAE. This was the ImageNet64 pre-trained model \cite{Chrabaszcz2017} whose architecture comprised 65 layers for the encoder and 75 for the decoder, mostly composed of convolutional and residual blocks. The GRU-based network, encoding the latent representation of the $\beta$ signal, was made up of basically two bidirectional GRU-based layers, followed by two fully-connected layers, totalling 12,400,344 parameters, and was obtained through empirical studies (Table \ref{tab:nn_architecture}). The GRU architecture was chosen due to its suitability for 1-dimensional input, enabling its temporal memory capabilities to be effectively adapted for handling spatial variations across different brain regions. The output was a set of latent variables that were then inserted into the decoder of the VDVAE to generate the first reconstruction. In a general case, being the VDVAE a hierarchical VAE, the generation process involves the sampling of some latent variables (each of length of 16) from a standard normal distribution in each layer of the decoder, such that the latent variables of the bottom layers are conditioned on the ones of the top layers (here, the top is intended as the bottleneck of the network, i.e. the conjunction between the encoder and the decoder, the top as the output layer). In our case, instead of sampling all the latent variables from the Gaussians, we used the GRU-based network to infer some of them directly from the beta weights. The latent variables to predict had to be chosen starting from the deep layers, being the following layers' latent variables conditioned on them. We choose the first 15 layers, i.e. a 13,334-length vector of latent variables, as a first option, and 31 layers, i.e. a 91,168-length vector, as a second option. At inference stance, the resulting 1-dimensional vector was hierarchically decomposed into individual latent variables, each of length 16. These latent variables were then fed into the corresponding top-down blocks of the VDVAE. Following a 1×1 convolution, they were integrated into the main decoding pathway of the decoder, where they functioned as conditioning factors guiding the generation process.
The GRU-based network model was selected as the optimal architecture following an ablation process. That was tested against CNN and transformer-based architectures exploiting the metrics described in the next paragraph (see par. \ref{subsec:qualitative_and_quantitative_evaluation_metrics}). 
The CNN-based architecture featured a bidirectional GRU layer, followed by layer normalization to enhance training stability. This was succeeded by a convolutional block consisting of two 1D convolutional layers with ReLU activations and max pooling, designed to extract hierarchical spatial features. A second bidirectional GRU layer, complemented by layer normalization, refined the spatial representations. The architecture concluded with a fully connected module incorporating batch normalization and dropout to translate the learned features into the final output.
The transformer-based architecture mirrored the CNN-based design, replacing the convolutional block with a transformer module. This module leveraged multi-head self-attention to model long-range dependencies and captured contextual relationships between features. Residual connections and layer normalization were incorporated to improve learning efficiency and maintain gradient flow.
\begin{figure}[t]
    \centering
    \includegraphics[width=0.95\linewidth]{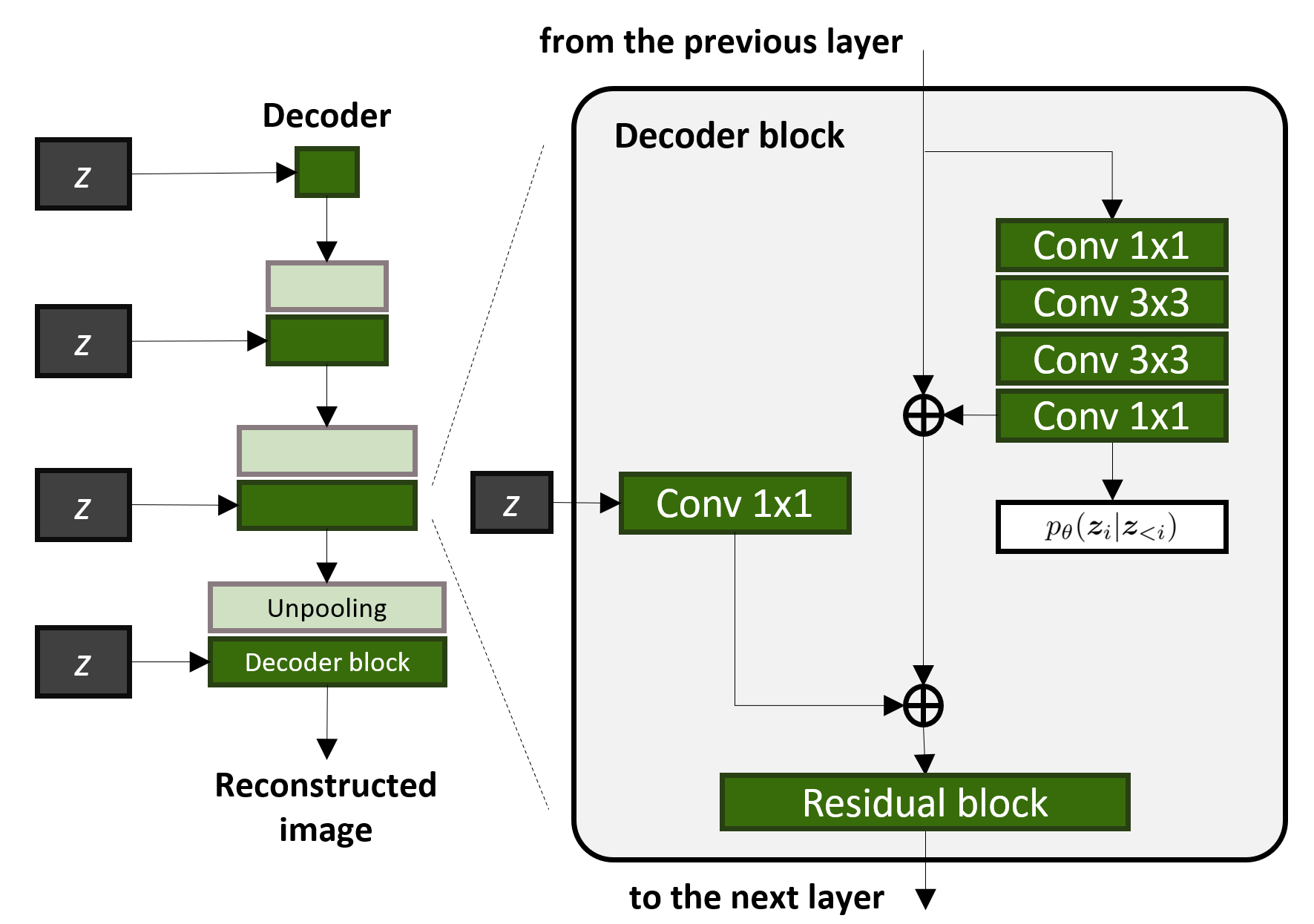}
    \caption{VDVAE decoder architecture. Latent variables $z$, predicted by the GRU-based architecture, were input into each decoder block through a 1 $\times$ 1 convolution.}
    \label{fig:DecoderVDVAE}
\end{figure}
\begin{table}[t]
    \centering
    \caption{Architecture of the proposed neural network to process in input the $\beta$ data and produce in output the encoding (latent representation). As an example, the input size was tailored for subject \#1.}
    \renewcommand{\arraystretch}{1.2}
    \begin{tabular}{l r}
        \hline
        Layer (type) & Output Shape \\ \hline
        Input & (*, 1, 15724)\\ \hline
        Bidirectional GRU & (*, 1, 200) \\ \hline
        Layer Normalization & (*, 1, 200) \\ \hline
        Dropout ($p=0.5$) & (*, 1, 200) \\ \hline
        Bidirectional GRU & (*, 1, 200) \\ \hline
        Layer Normalization & (*, 1, 200) \\ \hline
        Fully Connected & (*, 200) \\ \hline
        Batch Normalization & (*, 200) \\ \hline
        ReLU & (*, 200) \\ \hline
        Dropout ($p=0.5$) & (*, 200) \\ \hline
        Fully Connected & (*, 13344) \\ \hline
    \end{tabular}
    \label{tab:nn_architecture}
\end{table}

\subsubsection{Second stage: LDM with beta-predicted CLIP embeddings}
The second stage utilized the so-called Versatile Diffusion model, a pre-trained variant of the classical LDM \cite{Xu2023}. In such a model,  the generation process was conditioned on CLIP embeddings, including both vision and text latent descriptors (Fig. \ref{fig:architecture}). These CLIP embeddings were predicted from the beta coefficients by training two ridge regression models. For CLIP-vision, the ground truth embeddings were derived from the actual images, while for CLIP-text, the embeddings came from the captions of corresponding images in the COCO dataset. Thus, these models took a 1-D, 15,724-length signal (for subject 1) as input and produced a 77$\times$768 (text) or 257$\times$768 (vision) CLIP embedding as output. According to \cite{Ozcelik2023}, both regressors achieved optimal results with $\alpha$ regularization factor equal to 50,000. Hence, the reconstruction process was semantically driven using high-level semantic information provided by the CLIP-text information.

\subsection{LDM: noise sensitivity analysis}
To test robustness of LDM against noise, we systematically distorted the initial guess provided by the first stage using five different setups. A Gaussian-noised image (mean: 0.0, std: 1.0) was generated and superimposed onto the guess with progressively increasing amplitudes of 8, 16, 32, 64, and 256. Values exceeding the range of 0 to 256 were clipped to the corresponding boundary values.

\subsection{Evaluation metrics}
\label{subsec:qualitative_and_quantitative_evaluation_metrics}
Along with traditional vector-related metrics, such as Mean Squared Error (MSE) and Mean Absolute Error (MAE), two image-related metrics were considered to quantify the similarity between ground truth and predicted stimuli. Both low- and high-level (semantic) similarity scores were computed in the image-related metrics class.

\subsubsection{Image-similarity metrics: low-level}
The first low-level evaluation metric is the Structural Similarity Index Metric (SSIM) \cite{Wang2004}. This metric measures the similarity between the original and predicted images by evaluating structural information within each image, including features like edges, contrast, and texture. A score of 1 indicates perfect similarity, whereas a score of 0 signifies no similarity. The SSIM formula is given by:

\begin{equation}
SSIM (d_{x}, d_{y}) = \\
\frac{1}{N}\sum_{i=1}^{N} \frac{(2 \mu_{{x}_{i}} 2\mu_{{y}_{i}}+C_{1}) (2 \sigma_{{x}_{i}{y}_{i}} +C_{2})}{(\mu_{{x}_{i}}^{2}+\mu_{{y}_{i}}^{2}+C_{1})(\sigma_{{x}_{i}}^{2}+\sigma_{{y}_{i}}^{2}+C_{2})}
\label{eq::ssim}
\end{equation}

\noindent where $\mu_{{x}_{i}}$ and $\mu_{{y}_{i}}$ are the average pixel values at the $i$-th step of the $x$ and $y$ sliding windows on the target $d_{x}$ and predicted image $d_{y}$ respectively, $\sigma_{{x}_{i}}$ and $\sigma_{{y}_{i}}$ are the corresponding standard deviations, $\sigma_{{x}_{i}y_{i}}$ represents the covariance of $x_{i}$ and $y_{i}$, $C_{1}=(k_{1}L)^{2}$ and $C_{2}=(k_{2}L)^{2}$ are constants used to stabilize the division with a weak denominator, $L=2^{\#\text{bits per pixel}}-1$ is the dynamic range of the pixel values, and $k_{1}=0.01$ and $k_{2}=0.03$ are default constants.
A second low-level metric is the pixel correlation, i.e. the Pearson product-moment correlation coefficient: a value of 1 means a perfect positive relationship between the two images, a value of -1 means a perfect negative relationship, while a value of 0 indicates no relationship.
Two additional metrics were considered, which make use of the AlexNet architecture, as described in \cite{Ozcelik2023}. Two images are inserted into an AlexNet architecture and the latent features produced by the second layer and the fifth layer are compared. The former is used to compute the AlexNet(2) metric, while the latter AlexNet(5) metric, respectively. The similarity is computed through a two-way comparison \cite{Rakhimberdina2021}.

\subsubsection{Image-similarity metrics: high-level} 
Two high-level metrics have a similar functioning as the AlexNet(2) and AlexNet(5) presented just before. These are the ones derived from InceptionV3 \cite{Szegedy2016} and CLIP, which are the 2-way comparison of the last pooling layer of InceptionV3, and the output of CLIP-vision respectively. Finally, EfficientNet-B \cite{Tan2019} and SwAV-ResNet50 \cite{Caron2020} models were used for the other two high-level metrics. The features coming from those architectures are compared using the Spatial Distance Correlation (SDC):
\begin{equation}
    \text{SDC}(x, y) = 1 - \frac{(x - \bar{x}) \cdot (y - \bar{y})} {{|| x - \bar{x} ||_2} {|| y - \bar{y} ||_2}}
    \label{eq:highlevelmetrics}
\end{equation}
where $x$ and $y$ are the two vectors to compare and $\bar{x}$ and $\bar{y}$ their average values.

\section{Experimental sessions and results}
The processing pipeline was implemented using Python (version 3.8.13) and Pytorch \cite{Paszke2017}. The training of the proposed GRU-based regressor \ref{subsec:first_step_VDVAE_reconstruction}) and all the tests were performed on an NVIDIA A100 GPU with 40GB of RAM.

\subsection{Model ablation for latent variable estimation}
The comparison over the three architectures tested (GRU-, Convolutional- and Transformer-based) for the estimation of the latent variables of the beta coefficients in the first stage was favorable to GRU-based model (Table \ref{tab:ablation_architectures}). The test, performed over the data of subject \#1 using 15 latent variables, provided the best MSE and MAE results, in both validation and test sets.

\begin{table} [t]
    \centering
    \caption{Quantitative comparison between the three different architectures predicting the 15 latent variables of the VDVAE for subject 1.}
    \label{tab:ablation_architectures}
    \begin{tabular}{c|cc|cc}
        \toprule
        & \multicolumn{2}{c}{\textbf{Validation}}  &  \multicolumn{2}{c}{\textbf{Test}} \\ 
         \midrule
        & MSE$\downarrow$ & MAE$\downarrow$ & MSE$\downarrow$ & MAE$\downarrow$ \\
         \midrule
       Convolutional & 0.0687 & 0.1392 & 0.0683 & 0.1389 \\
       Transformer &  0.0690 & 0.1392 & 0.0685 & 0.1389 \\
       GRU &  \textbf{0.0685} &  \textbf{0.1390} & \textbf{0.0681} &  \textbf{0.1388} \\
         \midrule
        \bottomrule
    \end{tabular}
\end{table}
\begin{table} [t]
    \centering
    \caption{Comparison between GRU-based network and linear ridge regressor \cite{Ozcelik2023}. 15 latent variables were considered as the output of the prediction.}
    \label{tab:comparison_VDVAE_reconstuction}
    \begin{tabular}{c|cc|cc}
        \toprule
        & \multicolumn{2}{c}{\textbf{Validation}}  &  \multicolumn{2}{c}{\textbf{Test}} \\ 
         \midrule
        & MSE$\downarrow$ & MAE$\downarrow$ & MSE$\downarrow$ & MAE$\downarrow$ \\
         \midrule
        Ridge regressor \cite{Ozcelik2023} & 0.0695 & \textbf{0.1184} & 0.0690 & \textbf{0.1182} \\
       Ours &  \textbf{0.0685} & 0.1390 & \textbf{0.0681} & 0.1388 \\
         \midrule
        \bottomrule
    \end{tabular}
\end{table}

\subsection{Reconstruction accuracy}
MSE and MAE scores (subject \#1), computed using 15 latent variables in the GRU-based model for the first stage, were compared to the results in \cite{Ozcelik2023} adopting ridge regression  (Table \ref{tab:comparison_VDVAE_reconstuction}). The proposed architecture achieved a lower MSE 0.0681 (against 0.0690). Conversely, a higher MAE was attained suggesting that the GRU-based model was better than state-of-the-art model in minimizing significant deviations.
The models were also evaluated across the entire pipeline, considering the final reconstruction after both the VDVAE and LDM processes. The GRU-based model, tested with both 15 and 31 latent variables, demonstrated superior SSIM performance for both configurations (see Table \ref{tab:comparison_full_reconstruction}). Notably, the 31-latent variable model outperformed the results reported by \cite{Ozcelik2023} in CLIP performance, achieving a score of 0.958. While EfficientNet-B results were lower, the SwAV-ResNet50 performed slightly better, and InceptionV3 showed comparable values. However, AlexNet appeared to negatively impact our architecture's performance in both the (2) and (5) cases. When considering all four subjects (1, 2, 5, 7), the GRU-based model's performance was comparable to that reported by \cite{Lin2024, Takagi2023, Gu2024, Ozcelik2023} (Table \ref{tab:comparison_full_reconstruction_4subj}), though it generally exhibited lower performance metrics then \cite{Ozcelik2023}. This was particularly evident in pixel correlation, AlexNet(2), and AlexNet(5) evaluations.

\begin{table} [t]
    \scriptsize
    \centering
    \caption{Quantitative comparison between our model and the model by  Ozcelik et al. in terms of reconstruction (after the full pipeline) performance on the test set, in various configurations, according to the type of architecture used for the fMRI-VDVAE latent variables regression and the number of the VDVAE latent variables used. These statistics were computed over subject \#1 only.}
    \label{tab:comparison_full_reconstruction}
    \begin{tabular}{cc|cccc|cccc}
        \toprule
        \multicolumn{2}{c}{\textbf{Architecture}} & \multicolumn{4}{c}{\textbf{Low-level metrics}}  &  \multicolumn{4}{c}{\textbf{High-level metrics (eq. \ref{eq:highlevelmetrics})}}\\ 
         \midrule
        & Latents & SSIM$\uparrow$ & PixCorr$\uparrow$ & AlexNet(2)$\uparrow$ & AlexNet(5)$\uparrow$ & InceptionV3$\uparrow$ & CLIP$\uparrow$ & EfficientNet-B$\downarrow$ & SwAV-ResNet50$\downarrow$ \\
         \midrule
        Ridge \cite{Ozcelik2023} & 15 & 0.358 & 0.286 & 0.935 & 0.962 & 0.881 & 0.925 & 0.772 & \textbf{0.413} \\
        Ridge \cite{Ozcelik2023} & 31 & 0.367 & \textbf{0.305} & \textbf{0.967} & \textbf{0.974} & 0.878 & 0.925 & \textbf{0.768} & 0.415 \\
        Ours (CLIP) & 15 & 0.357 &  0.230 & 0.870 & 0.922 & 0.860 & 0.916 & 0.778 & 0.437 \\
        Ours & 15 & 0.370 &  0.242 & 0.904 & 0.942 & 0.868 & 0.924 & 0.779 & 0.421 \\
        Ours & 31 & \textbf{0.374} &  0.207 & 0.880 & 0.929 & \textbf{0.890} & \textbf{0.958} & \textbf{0.768} & 0.416 \\
         \midrule
        \bottomrule
    \end{tabular}
\end{table}

\begin{table} [t]
    \scriptsize
    \centering
    \caption{Quantitative comparison between our and state-of-the-art models \cite{Lin2024, Takagi2023, Gu2024, Ozcelik2023} in terms of reconstruction (after the full pipeline) performance on the test set, in various configurations, according to the type of architecture used for the fMRI-VDVAE latent variables regression and the number of the VDVAE latent variables used. In this case, 4 subjects (\#1, \#2, \#5, \#7) were considered and metrics values were averaged.}
    \label{tab:comparison_full_reconstruction_4subj}
    \begin{tabular}{cc|cccc|cccc}
        \toprule
        \multicolumn{2}{c}{\textbf{Architecture}} & \multicolumn{4}{c}{\textbf{Low-level metrics}}  &  \multicolumn{4}{c}{\textbf{High-level metrics (eq. \ref{eq:highlevelmetrics})}}\\ 
         \midrule
        & Latents & SSIM$\uparrow$ & PixCorr$\uparrow$ & Alexnet(2)$\uparrow$ & AlexNet(5)$\uparrow$ & InceptionV3$\uparrow$ & CLIP$\uparrow$ & EfficientNet-B$\downarrow$ & SwAV-ResNet50$\downarrow$ \\
         \midrule
        Lin et al. \cite{Lin2024} & - & - & - & - & - & 0.782 & - & - & -  \\ 
Takagi et al. \cite{Takagi2023} & - & - & - & 0.830 & 0.830 & 0.760 & 0.770 & - & - \\ 
Gu et al. \cite{Gu2024} & - & 0.150 & 0.325 & - & - & - & - & 0.862 & 0.465 \\ 

        Ridge \cite{Ozcelik2023} & 31 & 0.356 & \textbf{0.254} & \textbf{0.942} & \textbf{0.962} & \textbf{0.872} & 0.915 & \textbf{0.775} & \textbf{0.423} \\
        Ours & 15 & \textbf{0.361} &  0.202 & 0.878 & 0.931 & 0.862 & 0.914 & 0.786 & 0.434 \\
        Ours & 31 & 0.360 &  0.177 & 0.864 & 0.923 & 0.869 & \textbf{0.922} & 0.785 & 0.434 \\
         \midrule
        \bottomrule
    \end{tabular}
\end{table}


\subsection{LDM: noise sensitivity analysis}
\begin{figure} [t]
    \centering
    \includegraphics[scale=0.47]{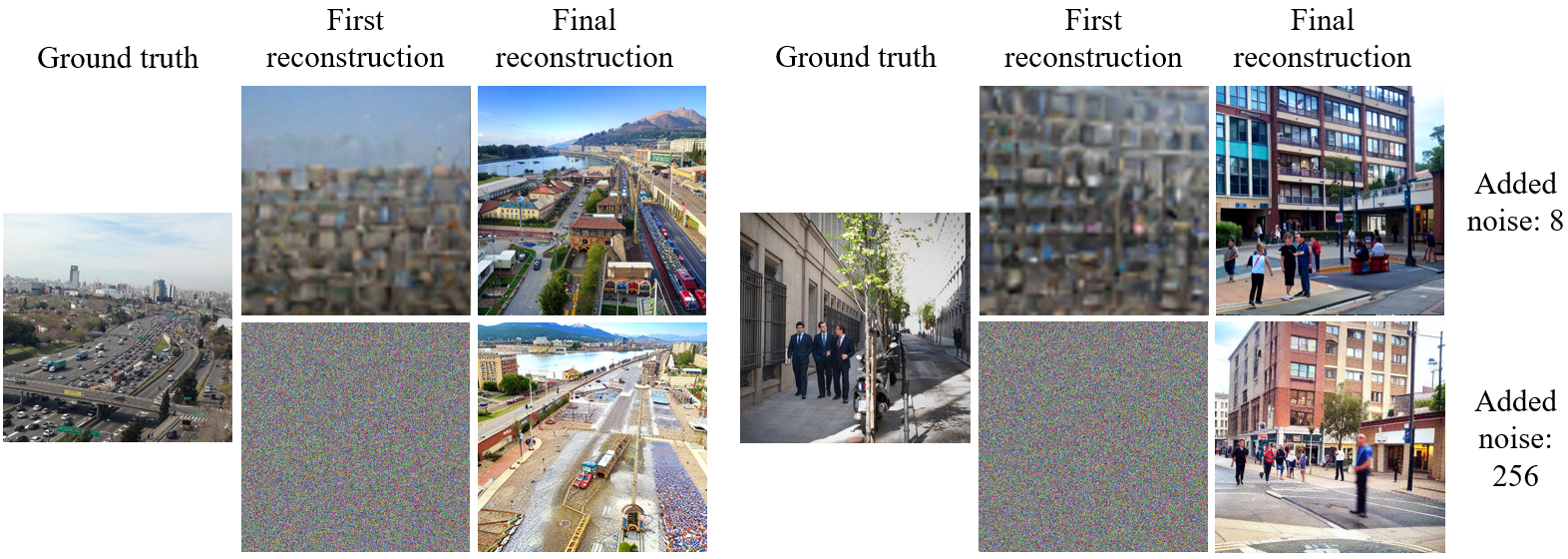}
    \caption{Two examples of reconstruction using 15 latent variables, when noise was applied to the first step reconstruction.}
    \label{fig:noise_example}
\end{figure}

We evaluated the effect of the first-stage reconstruction on the full processing pipeline for the 15-latent variable model applied to Subject \#1 by introducing noise into the initial reconstruction stage (Fig. \ref{fig:noise_example}). As expected the reconstruction quality for low-level metrics was directly related to the amount of noise (Table \ref{tab:ablation_noise}). Focusing on SSIM adding a Gaussian noise amplitude of 8 led to a decrease of the similarity of about 14\% with respect to the reference guess stimulus (no noise). Progressively increasing the noise led to a reduction of 20, 30, 42, and 65\%, respectively. Likewise, the pixel correlation metrics decreased as the noise increased up to 55\% for the widest noise amplitude. Conversely, the AlexNet-based metrics were less useful in assessing the noise effect in the predicted stimulus by the LDM.
High-level metrics revealed different trends. For the InceptionV3 metric, the model starting with maximum noise (256) performed better, suggesting that semantic content (e.g., CLIP) became more significant in the absence of a very coarse reconstruction. Likewise, the CLIP and EfficientNet-B metrics performed better as soon as the initial reconstruction was the noisiest. Finally, for SwAV-ResNet50, the trend was unclear so that the effect of noise appeared almost negligible as the metrics was 0.434 and 0.435 for no and maximum noise.

\begin{table} [t]
    \scriptsize
    \centering
    \caption{Quantitative comparison between different levels of noise given as input to the LDM as first stage reconstruction considering as the first reconstruction the one performed using 15 latent variables. The numbers 8, 16, 32, and 64 were the maximum values that can be summed to each pixel of the first reconstruction. 256 was the pure noise case in which a pure Gaussian noise was given as input to the LDM (hence, no first step was done).}
    \label{tab:ablation_noise}
    \begin{tabular}{c|cccc|cccc}
        \toprule
        \multicolumn{1}{c}{\textbf{}}  &
        \multicolumn{4}{c}{\textbf{Low-level metrics}}  &  \multicolumn{4}{c}{\textbf{High-level metrics (eq. \ref{eq:highlevelmetrics})}}\\ 
         \midrule
        & SSIM$\uparrow$ & Pixel correlation$\uparrow$ & AlexNet(2)$\uparrow$ & AlexNet(5)$\uparrow$ & InceptionV3$\uparrow$ & CLIP$\uparrow$ & EfficientNet-B$\downarrow$ & SwAV-ResNet50$\downarrow$ \\
         \midrule
        No noise & \textbf{0.361} &  \textbf{0.202} & 0.878 & 0.931 & 0.862 & 0.914 & 0.786 & 0.434 \\
        Added noise (8) & 0.311 & 0.197 & \textbf{0.907} & 0.948 & 0.879 & 0.946 & 0.767 & \textbf{0.400} \\
        Added noise (16) &  0.287 &  0.191 & 0.903 & 0.946 & 0.870 & 0.946 & 0.773 & 0.401 \\
        Added noise (32) & 0.256 & 0.180 & 0.894 & 0.944 & 0.866 & 0.949 & 0.759 & 0.409 \\
        Added noise (64) & 0.207 &  0.167 & 0.864 & \textbf{0.950} & 0.877 & \textbf{0.954} & \textbf{0.755} & 0.410 \\
        Added noise (256) & 0.140 & 0.090 & 0.783 & 0.924 & \textbf{0.907} & 0.939 & 0.766 & 0.435 \\  
         \midrule
        \bottomrule
    \end{tabular}
\end{table}

\subsection{Qualitative evaluation}
\begin{figure} [t]
    \centering
    \includegraphics[scale=0.75]{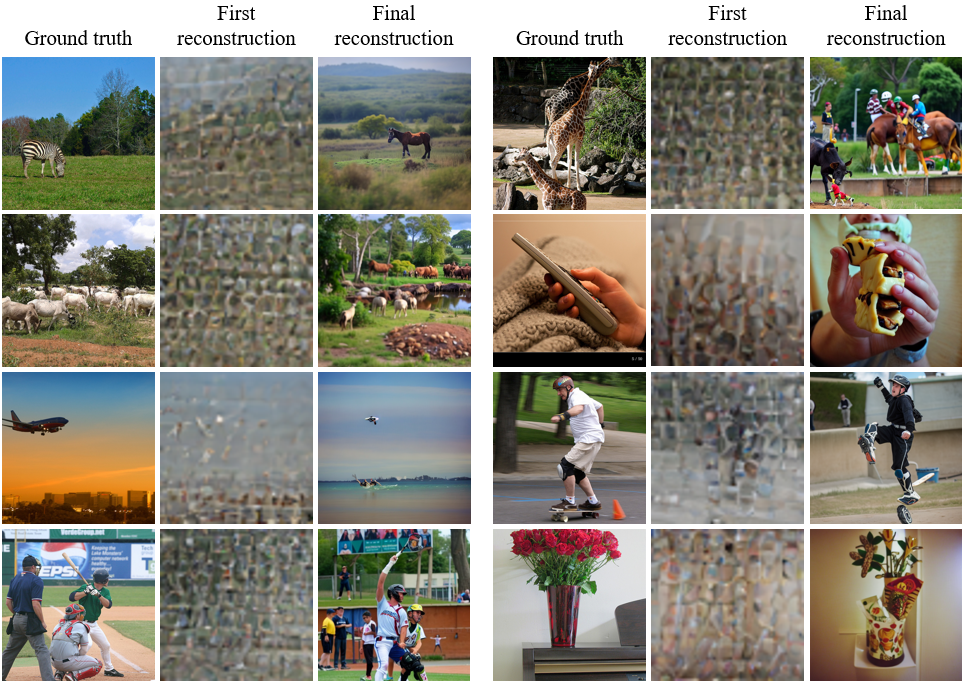}
    \caption{Examples of reconstructions in the 15 latent variables case. The ground truth of the test set was compared against the first-stage reconstruction and the second (final) reconstruction.}
    \label{fig:examples}
\end{figure}

Some examples of reconstruction using the 15-latent GRU-based model, distinguishing between the first and second steps were depicted, in comparison with the corresponding ground truth stimuli (Fig. \ref{fig:examples}). The first reconstruction appeared very blurry and vague, this due to the training of 15 layers of latent variables over the  77 layers in the VDVAE decoder. Nonetheless, the LDM was effective in reconstructing a visual context conveying significant semantic information starting from the blurred initial guess. The zebra on the grass was accurately predicted in a similar landscape alongside a horse. In the second stimulus, a herd of oxen in a woodland was reconstructed into a similar scene with different animals grazing. The process for the third stimulus (a commercial flight) seemed to confound the meaning, predicting a marine scene with a kite surfer. However, the primary concept of "flying" was properly identified. Nicely, the neural decoding accurately interpreted the baseball scene as an infield sport, depicting a scene with team members and coaches in the reconstructed image.

\section{Discussion} 
Experiments showed that modeling beta weights into a hierarchical latent space using GRU-based network, instead of a ridge linear model, (cfr. Fig. \ref{fig:architecture}), improved the visual reconstruction quality for subject 1 data by 2\% (SSIM). Likewise, the reconstructed image's semantics, measured by perceptual similarity, improved of about 4\% (crf. Table \ref{tab:comparison_full_reconstruction}). The results on the four subjects confirmed the quality of the reconstruction with respect to CLIP metrics (cfr. Fig. \ref{tab:comparison_full_reconstruction_4subj}). The noise sensitivity analysis of the LDM showed that the role of the first stage was fundamental to predict a stimulus featuring high structural similarity (crf. Table \ref{tab:ablation_noise}). Conversely, providing a large noise stimulus affected less the semantics of the predicted stimulus, while the structural similarity between the ground truth and predicted stimulus was very poor. These two main findings underscored the importance of leveraging non-linear relationships between BOLD space and the latent representation and two-stage generative AI for optimizing the fidelity of reconstructed visual stimuli from noisy fMRI data. 
Technically, we performed an ablation study on the neural architecture leading to identifying the GRU-based model against both convolutional- and transformer-based models (cfr. Table \ref{tab:ablation_architectures}). It was also found that using 31 latent variables to encode the beta values was unnecessary. Employing 15 (when using the proposed GRU-based network as a building block of the VDVAE decoder) was effective as well, featuring very similar SSIM and CLIP scores while offering the advantage of reduced computational cost (cfr. Table \ref{tab:comparison_full_reconstruction}).
Qualitatively, we showcased examples of image reconstructions using the 15-latent GRU-based model, illustrating the progression from initial to refined stages compared to ground truth stimuli. The initial reconstructions were notably blurry and indistinct, primarily due to training with a reduced number of latent variables (15 layers) compared to the 75 layers in the VDVAE decoder. Despite this initial fuzziness, the LDM demonstrated its effectiveness in reconstructing coherent visual scenes that conveyed significant semantic content. Interestingly, even when the prediction featured low similarity with respect to the nominal stimulus (cfr. Fig. \ref{fig:examples}, third example), the semantic coherence was kept grasping high-level contextual elements. Generally speaking, reconstruction inaccuracy may be split in two different categories, namely structural and semantic errors. Structural errors involved inaccuracies in the spatial arrangement or geometry of the reconstructed images. For instance, we noticed distortions in the shape of objects, such as elongated or compressed forms, which were not coherent with the general proportions of those classes of objects or did not correspond to the real stimulus. Semantic errors were caused by incorrect predictions of the CLIP embeddings, leading to mismatches in high-level features. They were particularly common when the classes have similar representations in the latent space.
Our architecture differed from those presented in \cite{Lin2024, Takagi2023, Gu2024} in several key components and decoding techniques. Specifically, \cite{Takagi2023, Lin2024} did not utilize a two-stage reconstruction process or employ any form of variational autoencoder for generating an initial rough reconstruction. Additionally, while \cite{Lin2024, Gu2024} used Generative Adversarial Networks (GANs) instead of Latent Diffusion Models (LDMs), \cite{Takagi2023} applied LDM as an adaptation of Stable Diffusion \cite{Rombach2022}. Our approach leveraged Versatile Diffusion \cite{Xu2023}, which was a more advanced diffusion model framework. Although CLIP-text embeddings were used across all these models for conditioning during the generation process, the use of CLIP-vision embeddings is unique in \cite{Ozcelik2023} and our work. Finally, the use of a non-linear technique such as GRU was novel to our knowledge.
From a physiological point of view, according to \cite{Wang2022}, our work contributed to support the idea that fMRI data describe neural representation of visual perception, featuring a relevant contribution of the visual cortex. According to \cite{Ozcelik2023}, we found that the CLIP embeddings, exploiting the textual encoding of the fMRI data, conveyed semantic information in the neural decoding process. Hence, multimodality had a key role in enhancing the performance of our decoding model. The predicted textual and visual semantic information from CLIP significantly guided the LDM during the generation process. This also validated the presence of semantic information in the ROIs we analyzed.
It is essential nonetheless to interpret the findings of this study within the context of its limitations. In detail, despite the CLIP metrics, the results across the four subjects of low-level metrics (cfr. Fig. \ref{tab:comparison_full_reconstruction_4subj}), especially SSIM, were inconclusive to support a generalization of the prediction in case of large signal to noise ratio as in the case of subject \#7. Moreover, only two sizes of the latent space representation were tested, namely 15 and 31. The first one was selected heuristically as a trade-off between computational effectiveness and gained performance. The last one was reported in \cite{Ozcelik2023} as the optimal trade-off between reconstruction performance and dimensionality. Concerning the model's complexity, we took significant steps to mitigate this issue by introducing a GRU-based decoder and reducing the number of latent variables. Further optimization will focus on developing a more lightweight LDM. As far as the dataset is concerned, we are fully aware that the study is limited to a specific dataset and a small number of subjects. Nonetheless, the NSD dataset includes a large number of images per subject that allowed to consistently test the intra-subject performance of the visual stimulus reconstruction. However, we remark that the access to 7T MRI scanners is really prohibitive, preventing a straightforward extension of the dataset requiring time consuming acquisition protocols.

\section{Conclusions}
We presented a novel approach to neural decoding by combining the power of deep neural networks and generative models. Specifically, we proposed a GRU-based neural network predicting latent variables of a VDVAE in the first processing stage. The second stage enhanced the reconstruction through LDM conditioned on CLIP embeddings derived from the fMRI data. This provided a high-resolution and semantically coherent image. The experimental results, based on the Natural Scenes Dataset (NSD), highlighted the effectiveness of our approach, consistently comparing to the recent literature. The reconstructed image not only captured low-level details, such as colors and shapes, but also kept high-level coherence with the original stimulus. This method proved crucial for enhancing the accuracy of reconstructed visual stimuli derived from noisy fMRI data. In summary, these findings emphasized the significance of utilizing non-linear relationships between BOLD signals and latent representations, as well as employing a two-stage generative AI approach. Once properly validated, such models might disclose new relevant knowledge about the human visual perception system.

\section*{CRediT authorship contribution statement}
\noindent \textbf{Lorenzo Veronese}: Conceptualization, Software, Data curation and analysis, Writing – original draft. \textbf{Andrea Moglia}: Investigation, Writing review. \textbf{Luca Mainardi}: Resources, Writing review. \textbf{Pietro Cerveri}: Conceptualization, Project administration, Funding acquisition, Supervision, Writing – review \& editing.

\section*{Declaration of interest statement}
\noindent The authors declare that they have no known competing financial interests or personal relationships that could have appeared to influence the work reported in this paper.


\section*{Acknowledgment}
\noindent This work was supported by P.E. PE00000013-FUTURE ARTIFICIAL INTELLIGENCE RESEARCH (FAIR), Italian Ministry of Research and University.

\bibliographystyle{plain}






\end{document}